\documentstyle[12pt]{article}
\def\Nc{$N_c\ $}

\def\dmud{{\nabla_\mu}}

\def\dmup{{\nabla^\mu}}

\def\undemi{{\scriptstyle1\over\scriptstyle2}}

\def\fpi{F_\pi}
\def\fk{F_K}
\def\mpi{M_\pi}
\def\mk{M_K}
\def\mhat{\hat m}
\def\fv {F_V}

\def\gs{G_S}
\def\gp{G_P}
\def\mv{M_V}

\def\ms{M_S}
\def\mp{M_P}
\def\yz{p_\alpha p_\beta}
\def\yu{q_\alpha q_\beta}
\def\yd{p_\alpha q_\beta}
\def\xu{q_\alpha p_\beta}
\def\xz{\eta_{\alpha \beta}}
\def\Vmud{V_\mu}
\def\Vmup{V^\mu}

\def\Vmunud{V_{\mu\nu}}
\def\Vmunup{V^{\mu\nu}}

\def\vmunup{v^{\mu\nu}}

\def\Pd{P^{(2)}_{\alpha\beta} }
\def\Pq{P^{(4)}_{\alpha\beta} }
\newcommand\pr[1]{ {\sl Phys. Rev.} {\bf #1} }
\newcommand\pl[1]{ {\sl Phys. Lett.} {\bf #1}}
\newcommand\prl[1]{{\sl Phys. Rev. Lett.} {\bf #1}}
\newcommand\np[1]{{\sl Nucl. Phys.} {\bf #1}}

\newcommand{\lbl}[1]{\label{eq:#1}}
\newcommand{\rf}[1]{(\ref{eq:#1})}
\newcommand{\vs}[1]{\rule[- #1 mm]{0mm}{#1 mm}}
\newcommand{\eq}{\vs{2}\begin{equation}}
\newcommand{\en}{\end{equation}}
\newcommand{\bea}{\begin{eqnarray}}
\newcommand{\ena}{\end{eqnarray}}
\def\cpt{$\chi{\rm pt}\ $}

\begin{document}
\rightline{IPNO/TH 94-26}
\bigskip

\centerline{\bf CHIRAL SYMMETRY ASPECTS OF THE SCALARS\footnote{
Talk given at the workshop "Two-photon physics from DA$\Phi$NE to
LEP200 and beyond", Paris 2-4 Feb. 1994} }
\bigskip
\centerline{B. MOUSSALLAM and J. STERN}
\smallskip
\centerline{\tensl Division de Physique
Th\'eorique\protect\footnote{Unit\'e de
Recherche des Universit\'es Paris 11 et Paris 6 associ\'ee au CNRS}
, Institut de Physique Nucl\'eaire}
\centerline{\tensl Universit\'e Paris-Sud, 91406, Orsay, France}
\bigskip
\centerline{ABSTRACT}
\smallskip
\begin{minipage}{5in}\baselineskip=12pt\tenrm
Electromagnetic decays of the scalar mesons are shown to be constrained
by chiral symmetry as a consequence of the fact that, in the chiral
limit, the two and three-point functions $<SS-PP>$ and $<VVS>$
satisfy super-convergent dispersion relations. The QCD asymptotic
behavior of the latter is canonical and it can be saturated by
a finite number of resonances.
The corresponding  chiral lagrangian for
vector and scalar resonances is constructed. Matching to the correct
asymptotic structure generates non-minimal terms which have so far
been ignored. It is found that the width $a_0(980)\to2\gamma$ can be
naturally reproduced suggesting that the $a_0(980)$ is not an exotic
particle.
\end{minipage}

\bigskip

\noindent{\large\bf 1- Introduction:}

The chiral symmetry limit of QCD, which is expected to be a
reasonably good approximation in the light quark sector, may be at the
origin of some of the surprising ( at first sight) properties that
the scalar mesons seem to have. We will concentrate here on the
question of the two photon width of the $a_0(980)$ (since this is
a gamma-gamma meeting) and it will be shown that the fact that this
width is anomalously small (as compared, for instance, with the
expectation from the naive quark model) has a natural explanation
in this limit, and is by no means incompatible with a standard
nonet status for the $a_0$.

Our starting point is to assume exact chiral symmetry (i.e.
we set $m_u=m_d=m_s=0$ in the lagrangian) and , furthermore, we will
work in the leading \Nc approximation. In these limits,
QCD still cannot be solved
in closed form, but analytic approximation methods, at least,
are available. The one that we will use is similar to the
boundary-layer technique used, for instance, in hydrodynamics and
consists in matching three asymptotic expansions for a given
correlation function: a) For large momenta,
the operator product expansion (OPE) generates an expansion in
inverse powers of the momenta  b)for small or moderate momenta the
correlation function is dominated by the contributions from the light
resonances and c)at very low momenta the expansion is generated
by chiral perturbation theory (\cpt).
The method is effective only for those n-point functions
which transform in a non trivial way under the chiral group and thereby
satisfy super-convergent dispersion relations.
Only in that case is it
possible to match the QCD asymptotic behavior with a finite number
of resonances.  This method has been used in the past for a diversity
of applications, for the first time by
Weinberg\cite{w67} who obtained an estimate for the
mass of the $a_1$ resonance.
Das et al\cite{das67} have evaluated in this way
the $\pi^+-\pi^0$ electromagnetic mass splitting. More recently, similar
ideas where discussed in the context of the effective lagrangian for
resonances\cite{eglpr} and an interesting application
to the question of whether the mass of the $u$ quark can be zero
can be found in ref.\cite{mu}.

\noindent{\large\bf 2- Meson spectroscopy:}

In the exact chiral limit, the spectrum of the lightest resonances
(mesons as well as baryons) in the QCD spectrum is expected to consist
of exactly degenerate octets (which, for the mesons, become nonets in
the leading \Nc approximation). This is not what is observed in practice,
of course, but already first order perturbation theory in terms of the
quark mass hamiltonian predicts a pattern of splitting, ideal mixing,
Gell-Mann-Okubo mass formulas, which is satisfied, in practice, to a
fairly good approximation. The fact that the pion nonet does not obey
ideal mixing is due to the axial anomaly and is
well understood in terms of the large \Nc expansion and
\cpt techniques (e.g. \cite{gl85} ). The only major exception to these
rules seem to be the scalar resonances.
The present candidates to form a nonet
that one can find in the review of particle properties\cite{rpp} are:
\bea
&I=1:\quad a_0(980),\quad\quad &I=1/2:\quad K_0^*(1430)\nonumber\\
&I=0:\quad f_0(975),\quad\quad &I=0  :\quad f_0(1400)
\ena
The decay properties of the $f_0(975)$ and the $f_0(1400)$ strongly
suggest that the flavor transformation properties of the former is
essentially like $s\bar s$ and the latter is $u\bar u+d\bar d$. The
pattern of masses then, clearly, is a complete mystery. Furthermore,
it has been argued recently that the $f_0(1400)$ should be renormalized
down to a very wide $f_0(1000)$\cite{morgan} which would almost
certainly imply that
$f_0(975)$ is an exotic. This interpretation requires that one identifies
the $s\bar s$ member of the nonet which should, if GMO is approximately
respected in this interpretation, lie somewhere in the 1.8 GeV range
(note that the $\theta(1700)$ is a possible candidate). At any rate,
it seems
important to assess whether the I=1 meson $a_0(980)$ is a member of the
nonet or an exotic particle. One of its properties, which was interpreted
as an indication for an exotic status
is the electromagnetic width\cite{rpp}(see \cite{mp} for a critical
discussion of the determination of these widths):
\eq
\Gamma(a_0(980)\to2\gamma)=(0.24\pm0.08)/Rate(\pi\eta)\quad{\rm KeV}
\en
which depends on the unknown $\pi\eta$ rate but, clearly, since this
is the only open decay channel, one does not expect it to be much smaller
than one even though some amount of $K\bar K$ decay is certainly
observed. In order to appreciate the peculiarity of this result, one
may compare it with the quark model calculation\cite{barnes}:
\eq
\Gamma(a_0\to2\gamma)\simeq 1.6\quad{\rm KeV}
\en
It is of theoretical interest to understand why the quark model fails
by so much in this case. We will argue that the amplitude for this
decay has to satisfy rather stringent chiral symmetry constraints
(certainly a surprising fact, at first sight, since no pion is involved
in that reaction) and this is ultimately why the naive quark model runs
into trouble.

\noindent{\large\bf 3- Chiral correlation functions:}

We intend to use the asymptotic matching technique for the three-point
function made out of two vector and one scalar current $<SVV>$ which
is obviously related to the decay amplitude of interest. In order
to determine some of the coupling constants we must discuss also the
two-point function $<SS-PP>$. Let us define these objects in a
more precise way. The vector, scalar and pseudoscalar currents, first,
are defined as follows:
\eq
j_\mu^a(x) =\bar\psi{\lambda^a\over2}\gamma_\mu\psi,\quad
j_S^a(x)=\bar\psi{\lambda^a\over2}\psi,\quad\quad
j_P^a(x)=i\bar\psi{\lambda^a\over2}\gamma^5\psi
\en
The couplings of these currents to the resonances which will play
a role in what follows are then defined  (in the chiral limit) as:
\bea
&<0\vert j_\mu^a(0)\vert V^b(p,\lambda)>=
\delta^{ab} M_V F_V\epsilon_\mu(\lambda),\quad &
<0\vert j_S^a(0)\vert S^b(p)>= \delta^{ab} M_S G_S\nonumber\\
&<0\vert j_P^a(0)\vert \pi^b(p)>= \delta^{ab} B_0 F_0\quad
&<0\vert j_P^a(0)\vert P^b(p)>= \delta^{ab} M_P G_P
\lbl{coupling}
\ena
Here, $F_0$ is the pion decay constant and $B_0=-<\bar u u>/F^2_0$, which
are standard notations in \cpt. Note that in the pseudo-scalar channel
we will take into account the pseudo-scalar resonance multiplet
($\pi'(1300)$) in addition to the pion multiplet, since it lies in
the same mass range as the vector or scalar resonances. Finally, the
definition of the two and three-point functions is as follows:
\bea
&&\Pi_{SP}^{ab}(p)=i\int \exp(ipx)<T( j_S^a(x)j_S^b(0) -
j_P^a(x)j_P^b(0) )>\nonumber\\
&&V^{abc}_{\alpha\beta}(p,q)=\int d^4x\,d^4y\exp(ipx+iqy) <T j^a_\alpha(x)
\,j^b_\beta(y)\, j^c_S(0) >
\lbl{vvs}
\ena
Let us begin by considering the two-point function $\Pi_{SP}$ in three
successive energy ranges:

\noindent a)$p\to\infty$:

Chiral symmetry implies that the operators which occur in the OPE
transform in the same way as the operator $SS-PP$. The important point is
that this operator transforms non-trivially under the chiral group
(like $(3,\bar3)+(\bar3,3)+...$ ). In addition there is a discrete
symmetry $\psi_L\to-\psi_L$, which implies that the operator of lowest
dimensionality in the OPE
must be of the kind $\bar\psi_L\psi_R\bar\psi_L\psi_R$.
Ignoring anomalous dimensions here, dimensional analysis shows that
for large scales:
\eq
\lim_{\lambda\to\infty}\Pi^{ab}_{SP}(\lambda p)\sim {1\over\lambda^4}
\en

\noindent b)$p\le 1-2$ GeV

In this region, we assume that the lowest lying set of resonances
dominate the two-point function. In the leading \Nc approximation
the only singularities are resonance poles so that, using
the definition of the coupling constants \rf{coupling}, one obtains
the following representation:
\eq
\Pi^{ab}_{SP}(p)= \delta^{ab}\left\{
{B^2_0F^2_0\over p^2}+{\mp^2\gp^2\over p^2-\mp^2}
-{\ms^2\gs^2\over p^2-\ms^2} \right\}
\en
Now in the very low energy region,

\noindent c)$p<< 1$ GeV:

Chiral perturbation theory
controls the expansion of the correlation functions. Up to one
loop, one obtains
\eq
\Pi^{ab}_{SP}(p)= \delta^{ab}F_0^2\left\{
{B^2_0\over p^2}-{5B^2_0\over96\pi^2F_0^2}\ln{-p^2\over\mu^2}+A_0(\mu)
+...\right\}
\en
We notice that, since $F_0^2$=O(\Nc), the second contribution which
is generated by the pion loop drops out in the leading \Nc
approximation and, in this limit, $A_0(\mu)$ is a scale independent
low-energy parameter. It is not difficult to determine this parameter
which controls, for instance, the expansion of the pseudo-Goldstone
boson masses in powers of the quark masses.  Up to and including
quadratic order, the expansion reads
\bea
&&\fpi^2\mpi^2=2\mhat B_0F_0^2+4\mhat^2A_0 F_0^2+...\nonumber\\
&&\fk^2\mk^2  =(r+1)\mhat B_0F_0^2 +(r+1)^2\mhat^2A_0F_0^2+...
\lbl{masses}
\ena
where
\eq
r={m_s\over\mhat}
\en
Note that at leading \Nc order loops are suppressed
and consequently, the expansion is analytic.
\cpt also allows one to determine the relation between $F_0$
and $\fpi$ to the same level of accuracy:
\eq
{\fpi^2\over F_0^2}=1+{2\over r-1}\left({\fk^2\over\fpi^2}-1\right)
\en
from which one can then solve for $\mhat B_0$ and $\mhat^2 A_0$ in terms
of the quark mass ratio $r$ and the well known quantities $\mpi$, $\mk$
$\fpi$, $\fk$. Now the value of $r$ is customarily assumed to be
$r\simeq26$, an estimate which may be obtained from \rf{masses} by
dropping the quadratic quark mass terms. However, an analysis of
the violation of the Goldberger-Treiman relation in the baryon
sector gives instead $6.3\le r\le 10$\cite{stern}. Here, we will assume
that $r$ is a free parameter that can vary in the whole range
between 6.3  and 26 and check, eventually, the sensitivity of our
results upon that.

Now, let us perform the matchings. Matching region b) with region c)
gives us one relation:
\eq
\gs^2-\gp^2=A_0F_0^2
\lbl{rel1}
\en
Then matching region b) with region a) gives a second relation:
\eq
\ms^2\gs^2=\mp^2\gp^2+B_0^2F_0^2
\lbl{rel2}
\en
which is exactly similar to Weinberg's first relation\cite{w67}.
Since we can calculate $A_0$ and $B_0$
(or rather $\mhat B_0$ and $\mhat^2 A_0$ which are RG invariant
combinations) as a function of $r$
from eqs.\rf{masses}, we can
now determine completely the two coupling constants $\gs$ and $\gp$
(note that only the RG invariant combinations $\mhat G$ actually occur
in physical quantities) which will be needed in the sequel. Before
we move to that let us show that, already, a non-trivial constraint
concerning the scalar meson spectrum emerges from the two relations
\rf{rel1}\rf{rel2}:

Relation \rf{rel2} implies that one can parametrize $\gs$ and $\gp$
in terms of an angle $\theta$
\eq
\ms\gs={B_0F_0\over\cos\theta}\qquad \mp\gp=B_0F_0\tan\theta
\en
Using the first relation we can solve for $\theta$ in the
following form:
\eq
\cos^2\theta={{\displaystyle \mp^2\over\displaystyle \ms^2}-1
\over{\displaystyle \mp^2\over\displaystyle  M_0^2}-1}\quad
{\rm where}\quad M^2_0={\mhat^2B_0^2\over\mhat^2A_0}
\en
Since $\cos^2\theta$ must be positive and smaller than one, $M_0$
$\mp$ and $\ms$ must satisfy one of the two inequalities:
\eq
\mp\ge\ms\ge M_0\quad{\rm or}\quad M_0\ge\ms\ge\mp
\en
Now $M_0$ is a function of $r$ which is easly seen to be increasing
and reaches $M_0=980$ MeV for $r=29.5$. This means that for any
reasonable value of $r$ the first inequality must hold i.e $\mp\ge\ms$.
Since $\mp\simeq M_{\pi'}=1.2-1.3$ GeV this inequality is compatible
with the assumption that $\ms=M_{a_0}=980$ MeV. On the contrary, it
would be violated if $a_0$ is assumed to be an exotic and the true
nonet mass were assumed to be 1.4 GeV.

Let us now turn to the VVS correlation function and, again, go
through the three energy regimes.

\noindent a)$p,q\to\infty$:

It is a simple matter to check that for $\lambda\to\infty$,
$V_{\alpha\beta}(\lambda p,\lambda q)$ scales as $1/\lambda^2$.
In addition, the leading term of the OPE is the scalar current
$\bar\psi\psi$ and, consequently, the corresponding Wilson coefficient
does not carry any anomalous dimension.
This means that
the asymptotic behavior is {\sl exactly} $1/\lambda^2$,
and one can as well impose matching to the full Lorentz structure
of the three-point function $V_{\alpha\beta}$. This
Lorentz structure can be found from the QCD tree graphs to be:
\bea
&&V^{abc}_{\alpha\beta}(p,q)_{p,q\to\infty}\sim
d^{abc}{B_0 F_0^2\over 4p^2 q^2 r^2}\,
\Big\{ (p^2-q^2+r^2)(p^2-q^2-r^2)\xz+4q^2\yz \nonumber\\
&&+4p^2\yu +2(p^2+q^2-r^2)\yd +2(p^2+q^2+r^2)\xu \Big\}(1+O
(\alpha_s)        )
\lbl{vvsasy}
\ena
where $r=-(p+q)$. Now in the resonance region:

\noindent b)$p,q \le 1-2$ GeV

We will include the poles corresponding to the lowest lying vector
and scalar nonets.
The $VVS$  Green's function is transverse $p^\alpha V_{\alpha\beta}=
q^\beta V_{\alpha\beta}=0$. It can be expressed
in terms of the two independent tensors with
this property:
\eq
P^{(2)}_{\alpha\beta}=p_\beta q_\alpha-p.q\,\eta_{\alpha\beta}\quad
{\rm and}\quad
P^{(4)}_{\alpha\beta}= q^2p_\alpha p_\beta +p^2q_\alpha q_\beta -
p.q\,p_\alpha q_\beta -p^2q^2\eta_{\alpha\beta}
\en
The most general form of the amplitude which matches with the
asymptotic behavior \rf{vvsasy} is the following:
\eq
V^{abc}_{\alpha\beta}={d^{abc}\over(p^2-\mv^2)(q^2-\mv^2)(r^2-\ms^2)}
\bigg\{ B_0 F_0^2\,\Big(\undemi(p^2+q^2+r^2)\Pd +\Pq\Big)+a\,\Pd
\bigg\}
\lbl{vvsres}
\en
One observes that the expression contains a {\sl single} arbitrary
parameter: $a$. As before, it can be determined by matching with
the low-energy regime:

\noindent c)$p,q\to 0$

where (forgetting about subleading pion loops) $VVS$ has the following
expansion:
\eq
V^{abd}_{\alpha\beta}(p,q)=-{9c\over5\mhat}d^{abc}(p_\beta q_\alpha
-p.q\eta_{\alpha\beta})+...
\en
One may recognize here the low-energy parameter $c$ which
contributes to the reaction $\gamma\gamma\to\pi^0\pi^0$ at chiral
order larger than four, which was recently studied in
\cite{bgs}( in this reference the parameter
is called $d_3$) and in \cite{kms}. A sum-rule was established for
$c$ in \cite{kms} which yields the following estimate
\eq
c={-(4.6\pm2.3) 10^{-3}\over r-1}
\lbl{c}
\en
The relation of the parameter $a$ to the low-energy constant $c$ is
easily determined by matching the low-energy region:
\eq
\mhat a={9\over5}\mv^4\ms^2 c
\lbl{ac}
\en
The preceeding discussion has a bearing on the effective lagrangian
description of the vector and scalar resonances, which is interesting
as such, and will provide an alternative derivation of the estimate
\rf{c} for $c$.

\noindent{\large\bf 4- Effective lagrangian for vector and scalar
resonances:}

We will denote the vector and scalar resonaces by $V_\mu$ and $S$
respectively and will assume that they both transform according to the
non-linear representation of the chiral group (see e.g.\cite{eglpr})
as $R\to h(\pi) Rh^\dagger(\pi)$. The effective lagrangian can be
used to compute the three-point function $VVS$ and it has to
reproduce the same one as found above \rf{vvsres}. This uniquely
determines the interaction part of the lagrangian with two vectors
and one scalar fields. Writing ${\cal L}_{eff} ={\cal L}_S
+{\cal L}_V+{\cal L}_{VVS}$, one has:
\bea
&&{\cal L}_S= \langle\dmud S\dmup S -\ms^2 S^2 -2\ms\gs S s\rangle\nonumber\\
&&{\cal L}_V=\langle-\undemi \Vmunud\Vmunup
+\mv^2\Vmud\Vmup+ f_V\Vmunud\vmunup\rangle
\ena
and
\bea
&&{\cal L}_{VVS}={1\over\fv^2\ms\gs} \langle{a\over\mv^2} (V_{\mu\nu} -
f_V v_{\mu\nu})^2 S\nonumber\\
&& -B_0F_0^2\Big[2\mv^2V_\mu V^\mu S - V_{\mu\nu}(V^{\mu\nu}
-f_V v^{\mu\nu})S
+{1\over2\mv^2}(V_{\mu\nu}-f_V v_{\mu\nu})^2\nabla^2 S\Big]\rangle
\lbl{lag}
\ena
where $f_V=\fv/\mv$.
(A form which is manifestly covariant is obtained by replacing the
vector source $v_{\mu\nu}$ by $1/2 f^+_{\mu\nu}$ and the
scalar source $s$ by $1/2\chi^+$). We will use this
effective lagrangian for two purposes. Firstly, if one is interested
in the decay $S\to2\gamma$ then there are two terms only
which contribute (in this formalism, the coupling of a photon to
a vector meson vanishes when the photon is on-shell):
\eq
{\cal L}_{S\gamma\gamma}={1\over\mv^4\ms\gs}\langle a Sv_{\mu\nu}v^{\mu\nu}-
{B_0F_0^2\over2}\nabla^2S v_{\mu\nu}v^{\mu\nu}\rangle
\lbl{sgg}
\en
This lagrangian differs from the one which was used e.g. in \cite{bgs}
by the presence of the second term, proportional to $B_0$.
This term is needed to enforce the QCD short distance behavior
even though it is non-leading at low energies. In a sense this is
an unsusual situation: up to now it seemed possible to achieve
compatibility of effective lagrangians for resonances with the
short distance properties of QCD by retaining only terms with a minimal
number of derivatives\cite{eglpr}. Here, we face a counterexample
to this situation. Both terms contribute to the decay $S\to2\gamma$
but only the minimal term contributes to the low-energy parameter
$c$ (or $d_3$). Notice, however, that this remark has no practical
bearing on the evaluation of the $\gamma\gamma\to\pi^0\pi^0$
cross-section in ref.\cite{bgs}

As a second application, the form of
${\cal L}_{eff}$ suggests an alternative way
to determine the parameter $a$. Let us set the scalar source equal to
the quark mass matrix: $s={\cal M}=diag(\mhat,\mhat,m_s)$.
The equation of
motion allows us to eliminate the scalar resonance field:
\eq
S=-{\gs\over\ms} {\cal M}
\en
Clearly, the lagrangian \rf{lag} then describes
propagation of vector mesons in
the presence of a vector source and the sole effect of the scalar
source is to induce ideal-mixing and mass-splitting in the vector
nonet. One observes  that the quark masses act on the kinetic energy
as well as on the vector mass term.
As a consequence, the actual meson masses get expressed as a function of
the chiral limit mass $\mv$ and the quark masses in the following way
\eq
M_\phi^2={\mv^2\over1+ 2m_sa/\fv^2\ms^2 +O(m_q^2) } , \quad
M_\rho^2=M_\omega^2={\mv^2\over1+ 2\mhat a/\fv^2\ms^2+O(m_q^2) }
\lbl{mvec}
\en
and
\eq
M_{K^*}^2={\mv^2\over1+ (\mhat+m_s) a/\fv^2\ms^2+O(m_q^2) }
\en
Note that the terms proportional to $B_0$ in the lagrangian
only contribute at order $O(m^2_q)$.
{}From \rf{mvec} one obtains an estimate of the parameter
$a$ in terms of the $\phi$ and $\rho$ meson masses.
\eq
\mhat a=\mv^4\ms^2\,{\fv^2\over2(r-1)}\Big({1\over m_\phi^2}-
{1\over m_\rho^2} \Big)
\lbl{cform}
\en
If one uses the $K^*$ and $\rho$ masses instead one obtains the same
result with a 10\% accuracy.
It turns out, in fact, that eq.\rf{cform} reproduces nearly exactly the
same expression for the constant $c$ (via \rf{ac}) as the one
found in
\cite{kms} by saturating the sum-rule with the contribution of
the vector mesons (in the approximation where $\fv^2=F_\rho^2=
F_\omega^2=F_\phi^2$ which can be shown to hold, using \rf{lag}
to first order in the quark masses). A remark is in order here.
At first order in the quark masses, an equally valid way to
determine $a$ would be to expand first the denominators in
\rf{mvec} and determine $a$ from the mass difference $M_\phi^2-
M_\rho^2$. The result is easily seen to differ from the
preceeding one by a factor $\simeq M_\phi^2/M_\rho^2$. This
essentially confirms the size of the error bar in the
determination of the constant $c$ in \rf{c}.

\noindent{\large\bf 5- Results and conclusions:}

Now that the unique parameter $a$ is determined it is not difficult,
using either directly the three-point function \rf{vvsres} or
the lagrangian form \rf{sgg}, to evaluate the amplitude for the decay
$S\to2\gamma$ and to calculate the width. The result is shown
in the figure as a function of the quark mass ratio $r$ and for
three values of the constant $c$, the central one and the two
extreme values allowed by the error bar in \rf{c}. One observes
that for the central value the width is in surprizingly good
agreement with experiment and depends very weakly on the value
of the quark mass ratio. Now if $c$ is allowed to be larger
(in magnitude) then the width is predicted to be also larger and
becomes more strongly dependent on $r$. The width can become as
large as 1 KeV if $r$ is large, $r=26$, and $c$ is at its upper
bound. In that case, the width can still be in qualitative agreement
with expermiment provided the value of the quark mass ratio is of the
order $r\simeq10$.

At this point we should also mention that there
is a further uncertainty in the calculation, which is associated
with the value of $\mp\simeq M_{\pi'}$ (in the figure
we have taken $M_P=1300$ GeV). Clearly, since the $\pi'$
is a rather wide resonance it is not a very precise approximation
to treat it in the narrow width limit. Despite all these uncertainties,
we feel that this calculation strongly suggests that there seems
to be nothing surprising about the small value of the electromagnetic
width of the $a_0$, which has been reproduced rather naturally.
Chiral symmetry was an essential ingredient which allowed
us to construct the correct form of the three-point function (or
the equivalent effective lagrangian) and to determine all the
parameters. The important role of chiral symmetry in this application
is at high energy rather than, as usual, at low energy.
This approach can be extended to discuss the main decay
mode of the $a_0$, $a_0\to\pi\eta$ (see \cite{ms} ). In that case,
the dependence upon $r$ is rather strong but it is not difficult
to reproduce the experimental result. All this seem to indicate that
there could be nothing exotic about the $a_0(980)$ (as e.g. in the
suggestion of ref\cite{cdgkr} ) which could well be an ordinary
nonet member.
An interesting question at this
point is whether the $f_0(975)$ is also a nonet member with an exotic
mixing angle or a completely exotic state. As has been pointed out
and discussed in \cite{close} this question
could be answered at Da$\phi$ne by
measuring the decay rates $\phi\to a_0\gamma$ and $\phi\to f_0\gamma$.
These rates can also be evaluated in our approach as a function
of the mixing angle, the results can be found in \cite{ms}.

\medskip
\noindent{\large\bf Acknowledgements:} We thank
F. Close, J. Gasser, M. Knecht and E. de Rafael for useful conversations.

\bigskip
\noindent{\large\bf Figure caption:} Result for the width $a_0\to2\gamma$ as
a function of the quark mass ratio for three values of the low-energy
parameter $c$ ($c=-10^{-3}\hat c/(r-1)$ ).

\end{document}